\def\prd{Phys. Rev. D}

\def\apj{Astrophys. J.}

\def\apjs{Astrophys. J.Suppl.}
\def\mnras{Mon. Not. R. Astr. Soc.}
\def\aap{Astr. Astrophys.}

\def\jcap{JCAP}

\def\procspie{Proc. SPIE}

\def \<{\langle}
\def \>{\rangle}

\newcommand{\ra}{\;\raise1.0pt\hbox{$'$}\hskip-6pt\partial\;}
\newcommand{\lo}{\;\overline{\raise1.0pt\hbox{$'$}\hskip-6pt\partial}\;}

\newcommand{\degree}{^\circ}

\newcommand{\Abs}{\abstract}
\newcommand{\Ack}{\acknowledgments}
\newcommand{\mktt}{\maketitle}

\documentclass[a4paper,11pt]{article}
\pdfoutput=1

\usepackage{jcappub,graphicx,epsfig,natbib,color,times,bm,amsmath,multirow,hyperref}
\usepackage[ddmmyyyy,hhmmss]{datetime}

\begin{document}

\title{From top-hat masking to smooth transitions: P-filter and its
application to polarized microwave sky maps}

\author[a,b]{Hao Liu,}\emailAdd{liuhao@nbi.dk}

\author[a]{James Creswell,}\emailAdd{creswelljames@gmail.com}

\author[a]{Sebastian {von Hausegger}}\emailAdd{s.vonhausegger@nbi.dk}

\author[a]{and Pavel Naselsky}\emailAdd{naselsky@nbi.dk}

\affiliation[a]{The Niels Bohr Institute \& Discovery Center, Blegdamsvej 17, DK-2100 Copenhagen, Denmark}

\affiliation[b]{Key Laboratory of Particle and Astrophysics, Institute of High Energy Physics, CAS, 19B YuQuan Road, Beijing, China}

\Abs{

In CMB science, the simplest idea to remove a contaminated sky region is to
multiply the sky map with a mask that is 0 for the contaminated region and 1
elsewhere, which is also called a top-hat masking. Although it is easy to use,
such top-hat masking is known to suffer from various leakage problems.
Therefore, we want to extend the top-hat masking to a series of semi-analytic
functions called the P-filters. Most importantly, the P-filters can seamlessly
realize the core idea of masking in CMB science, and, meanwhile, guarantee
continuity up to the first derivative everywhere. The P-filters can
significantly reduce many leakage problems without additional cost, including
the leakages due to low-, high-, and band-pass filtering, and the E-to-E,
B-to-B, B-to-E, and E-to-B leakages. The workings of the P-filter are
illustrated by using the WMAP and Planck polarization sky maps. By comparison
to the corresponding WMAP/Planck masks, we show that the P-filter performs
much better than top-hat masking, and meanwhile, has the potential to
supersede the principal idea of masking in CMB science. Compared to mask
apodization, the P-filter is ``outward'', that tends to make proper use of the
region that was marked as 0; whereas apodization is ``inward'', that always
kills more signal in the region marked as 1.

}


\mktt

\section{Introduction}\label{sec:intro of method}

During the last several years, analysis of the Cosmic Microwave Background
(CMB) polarization became one of the primary occupations for the
Planck~\citep{2014A&A...571A...1P, 2016A&A...594A...1P, 2018arXiv180706205P}
and forthcoming space/ground based experiments~\citep{2012SPIE.8442E..19H,
2016arXiv161002743A, 2011arXiv1110.2101K, quijote2012}. The polarized CMB
signal is usually characterized in terms of the Stokes parameters Q and U and
contains unique information about the properties of the early Universe,
especially cosmological gravitational waves (CGW).

Standard CMB analysis techniques involve transformation of the Q and U Stokes
parameters into scalar (E) and pseudo-scalar (B)
modes~\citep{PhysRevD.55.1830, 0004-637X-503-1-1}, where the B-mode is due
only to CGW, lensing effects, and residuals from foregrounds. The contribution
of the CGW into the CMB B-mode is characterized by the parameter $r$, the
ratio of tensor-to-scalar modes, where the term ``tensor'' corresponds to the
power spectrum of the cosmological gravitational waves, and ``scalar''
describes the power spectrum of primordial adiabatic perturbations, which are
responsible for the formation the galaxies and Large Scale Structure of matter
in an expanding Universe. The most stringent constraint, given by the
BICEP2-KECK/Planck joint data analysis, is
$r<0.07$~\citep{2015PhRvL.114j1301B}. Bearing in mind the target of
constraining $r$ to $10^{-4}$$\sim$$10^{-3}$, we should not only significantly
increase the sensitivity of experiments but also understand the properties of
the polarized foregrounds and systematics at least by 3-4 orders of magnitude
better than now. At the same time we need to revise some aspects of data
processing, like masking, band passing, apodization etc., which could be the
potential sources of uncertainty for extraction of the B-mode of polarization
at the level of $r\sim 10^{-4}$. In this work, we focus on improving the
masking technique.

In CMB science, the basic and most important idea to determine the shape of a
mask can be described by just one sentence: the mask should cover the
brightest pixels because they are contaminated by the foreground. This idea
was explicitly adopted by the WMAP team in~\cite{2003ApJS..148...97B}, where
they use temperature thresholds to define the mask regions; it was also
adopted by the Planck team, as we shall see in Section~\ref{sub:example of
p-filter}. Normally, further adjustment of the region definition is needed to,
e.g., make the edge flatter, reduce the number of tiny holes, block missing
pixels, cover the regions with strong systematics, etc.

Once the region is defined, masking is normally done by keeping the accepted
region unchanged and zeroing the rest, which is called top-hat masking. The
mask's profile is evidently non-analytic at the boundary. For temperature
anisotropy such non-analyticity is less critical in comparison with the case
of polarization, where the discontinuity of the first derivative at the
boundary plays an important role in mixing the E and B components. To minimize
this effect, it was proposed by~\cite{2006PhRvD..74h3002S} to smooth the mask
boundary and restore the analyticity of the mask. However, there are many
smoothing schemes and an even larger number of parameters, leaving the optimal
choice not at all obvious.

Another problem of top-hat masking is that it indiscriminately kills all
pixels in the rejected region. However, those pixels certainly have different
levels of foreground contamination, thus it is questionable to kill them all.
This problem can not be solved by ordinary ways of smoothing the mask, because
they are done after the mask was produced. However, it can be perfectly solved
by the P-filter method introduced in this work.

In this paper, the P-filter is designed as a natural extension of top-hat
masking that seamlessly implements the most important idea of masking in CMB
science. The P-filter can be seen as a family of semi-analytic functions
defined with two parameters: the threshold $P_t$, which implements the
principal idea of masking; and the rank parameter $a$, which determines the
shape of the filter. In the pixel domain, the P-filter essentially weights the
pixels above and below a given threshold differently. The shape of the filter
is automatically compatible with the actual map, suppressing the brightest
pixels, while preserving continuity of the first derivative everywhere. It is
also much easier to use the P-filter with maps having varying resolutions:
changing the resolution of a top-hat mask is not necessarily straightforward,
but a P-filter can be conveniently regenerated in a new resolution using
identical parameters, with all preferable features preserved.

Throughout this paper, unless mentioned elsewhere, we will use the WMAP 9-year
and Planck 2018 maps with $N_{side}=512$ smoothed to an angular scale of
$1\degree$. The outline of the paper is the following: in
Section~\ref{sec:filter definition and example} we introduce the mathematical
definition of the P-filter family and provide illustrating examples, and in
Section~\ref{sec:how to make a common P-filter}, we explain how to construct a
common P-filter for varies frequency bands. In Section~\ref{sec:what can
pfilter do for us} we explore a range of advantages P-filters have, and in
Section~\ref{sec:EB-leakage and optimization} we show example of how to
optimize the filter parameters based on a specific requirement. Finally, a
brief discussion is given in Section~\ref{sec:discussion}.

\section{Definition of the P-filter family and examples}\label{sec:filter definition and example}

\subsection{Definition of the P-filter}\label{sub:filter definition} 

The P-filter family proposed here consists of a set of filters defined as
follows (bold letters stand for sky maps):
\begin{eqnarray}\label{equ:p filter complete}
\mathbf{k}(P_t) & = & 
\left\{
\begin{array}{lr} 
1 & (\mathbf{P}\le P_t) \\ \nonumber
\left[1+\log\left(\frac{\mathbf{P}}{P_t}\right)\right] \frac{P_t}{\mathbf{P}} &
(\mathbf{P} > P_t)
\end{array}
\right. 
\\
\mathbf{k}(a,P_t) &=& \mathbf{k}(P_t)^a \\ \nonumber
\mathbf{P}' &=& \mathbf{k}(a,P_t)\cdot \mathbf{P},
\end{eqnarray}
where $P_t$ is the threshold, $a$ is the filter rank, $\mathbf{P}$ is the
input polarization intensity map, and $\mathbf{P'}$ is the filtered map.
$\mathbf{k}(P_t)$ represents the simplest form of the filter kernel with rank
$a=1$, and $\mathbf{k}(a,P_t)$ covers all possible shapes of the filter
kernels in terms of weight maps with the same resolution as $\mathbf{P}$.
\begin{figure}[!tbh]
  \centering
  \includegraphics[width=0.96\textwidth]{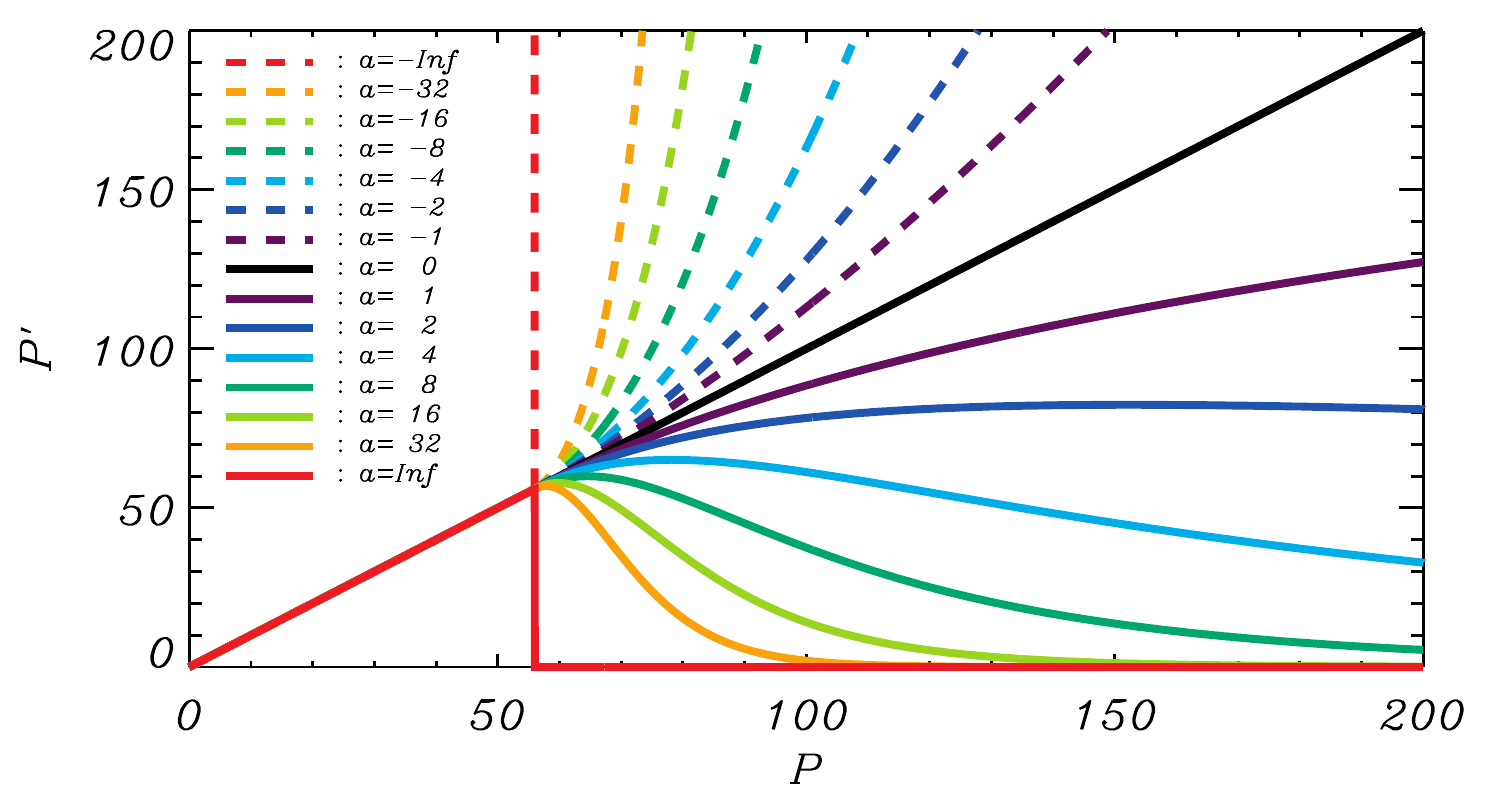}
  \caption{ Profiles of P-filters from Eq.~\ref{equ:p filter complete}, with
  various ranks $a$ and the arbitrary choice of $P_t\equiv56$ $\mu$K as
  example. The horizontal axis is the input polarization intensity, and the
  vertical axis is the filtered polarization intensity. $a=0$ corresponds to
  ``no filtering'', and $a=\infty$ is top-hat masking. }
  \label{fig:p-filter}
\end{figure}

According to the definition, the sky region with $\mathbf{P}<P_t$ will not
change, while other regions will be suppressed by the factor
$\mathbf{k}(a,P_t)$. Note that the P-filters are defined directly on
$\mathbf{P}$, where $\mathbf{P}$ can be any kind of data, even , e.g., time
series data. This means the P-filters potentially have much wider application
than merely in CMB science.

The first derivative of the filter function for $\mathbf{P}\ge P_t$ is:
\begin{align}\label{equ:p_filter 1st deri complete}
\frac{d P' }{d P} = 
\left(\frac{P_t}{P}\right)^a
\left(1+\log\frac{P}{P_t}\right)^{a -1} \left[1+(1-a
)\log\frac{P}{P_t}\right].
\end{align} 
For any finite rank $a$, Eq.~\ref{equ:p_filter 1st deri complete} is always
equal to 1 at $P_t$, no matter calculated from the left or the right side.
Thus the filter function is always continuous to the first derivative.
However, discontinuity of the filter does occur for the second derivative,
thus it is of differentiability class $C^1$ and we call it semi-analytic.
Continuity of the whole family up to the first derivative everywhere is a
desirable feature, e.g., with special design, it helps to prevent the
EB-leakage~\citep{2006PhRvD..74h3002S}.

In Figure~\ref{fig:p-filter}, we illustrate the shapes of the P-filter
functions when the rank $a$ varies from $-\infty$ to $+\infty$. From this
figure, it is easy to see how the P-filter family covers the entire region
where $\mathbf{P}>P_t$. Each point $(P,P')$ in this region belongs to
\emph{one and only one} P-filter. Since the filters with $a<0$ will make
$P'>P$ (the dashed lines), which is normally unwanted, in this work we
consider only those with $a\ge 0$, which means to suppress high intensities.
Some specific ranks are noteworthy, for example:
\begin{enumerate}
\item $a =0$ gives $\mathbf{P}'\equiv \mathbf{P}$ (no filtering).
\item $a =1$ gives the simplest P-filter.
\item $a =\infty$ gives top-hat masking.
\end{enumerate}

From Eq.~\ref{equ:p_filter 1st deri complete}, one can derive the conclusion
that $a =1$ is the maximal rank for which the P-filter monotonically increases
until $P=+\infty$. Any $a>1$ will lead to a point $P_1<+\infty$ where
$dP'/dP=0$ (see Figure~\ref{fig:p-filter}). The value of $P_1$ can be easily
calculated from Eq.~\ref{equ:p_filter 1st deri complete} as:
\begin{equation}\label{equ:turn point of 1st derivative}
P_1=P_t e^{\frac{1}{a -1}},
\end{equation}
thus, for a higher rank $a$, $P_1$ will be closer to $P_t$, and the filter
approaches a top-hat mask.

In practice, it is unnecessary to require monotonicity for all positive $P$.
Each map has a maximum value of the polarization intensity as $P_{max}$, thus
a monotonicity up to $P_{max}$ is practically sufficient. This enables us to
determine an upper limit for $a$ by solving $P_1=P_{max}$, which gives
\begin{equation}\label{equ:recomm rank}
a_{max} =  1+\frac{1}{\ln\frac{P_{max}}{P_{t}}}.
\end{equation}
This is an important way to reduce the number of parameters of the P-filter,
and to make it as simple as possible. If one accept $a_{max}$ given by
Eq.~\ref{equ:recomm rank}, and assumes $P_{max}/P_t=u$, then the point of
maximum for the filtered  polarization intensity is
\begin{equation}\label{equ:max p filtered with reom rank}
\frac{P'_{max}}{P_t} = u \left(\frac{1+\ln (u)}{u}\right)^{1+1/\ln (u)}.
\end{equation}

\subsection{Examples}\label{sub:example of p-filter}

\begin{figure}[!tbh]
  \centering
  \includegraphics[width=0.32\textwidth]{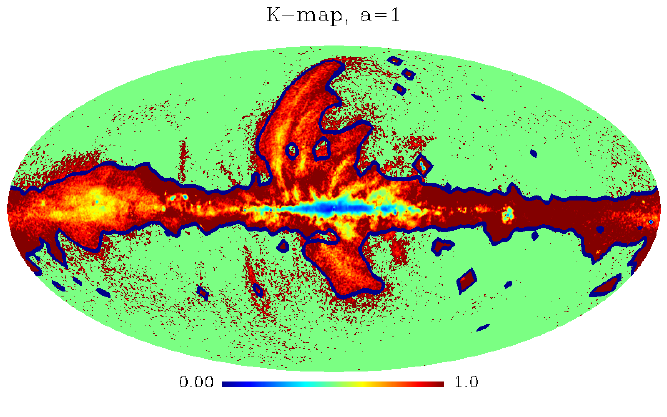}
  \includegraphics[width=0.32\textwidth]{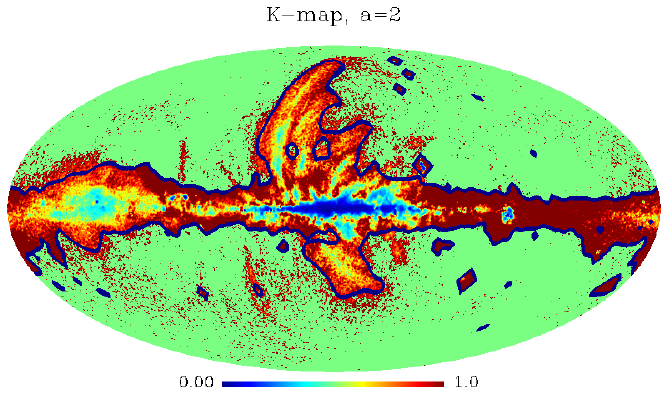}
  \includegraphics[width=0.32\textwidth]{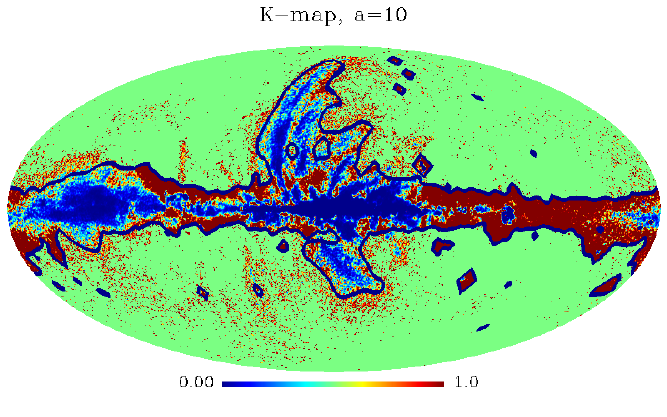}

  \includegraphics[width=0.32\textwidth]{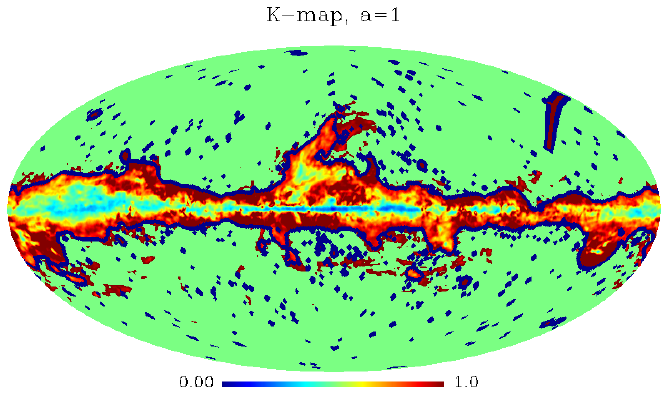}  
  \includegraphics[width=0.32\textwidth]{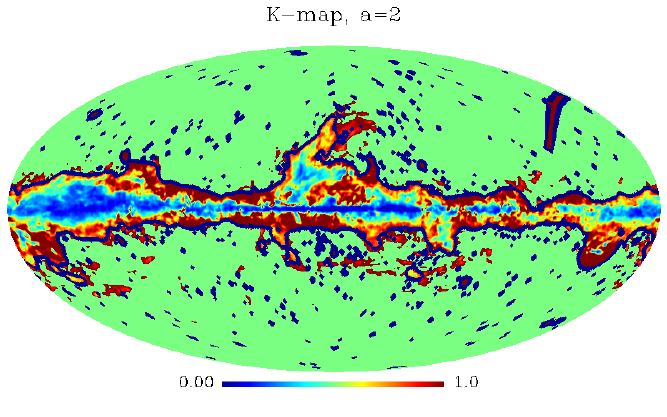}
  \includegraphics[width=0.32\textwidth]{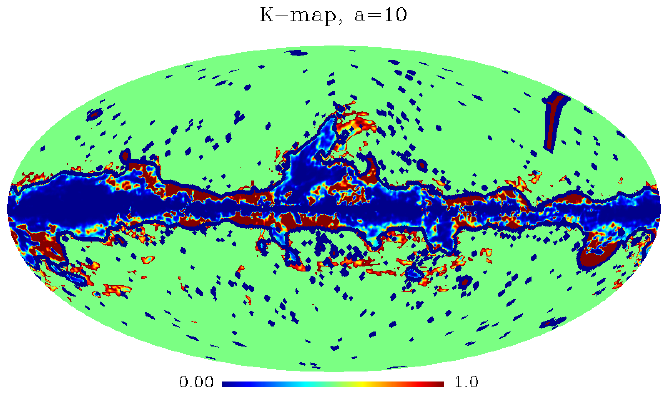}
  \caption{\emph{Upper}: the $\mathbf{k}(a,P_t)$ maps in comparison to PAM
   (dark blue contour line), where $\mathbf{k}(a,P_t)$ is generated from the
   WMAP 23 GHz band polarization intensity map with the same sky fraction
   ($\mathbf{k}=1$) as PAM, and the ranks are 1, 2 and 10 (from left to
   right). \emph{Lower}: similar to the upper panels but for the Planck 353
   GHz band polarization intensity map and CSM. Note that for better
   visibility, the area equal to 1 in both the $\mathbf{k}(a,P_t)$ map and the
   PAM/CSM masks is set to green.}
  \label{fig:p-filter1}
\end{figure}

As explained in Section~\ref{sec:intro of method}, the threshold $P_t$
implements the basic and most important idea of masking in CMB science:
masking out the brightest pixels. To show how this idea works with top-hat
masking and with the P-filter, we choose the WMAP 23 GHz polarization
intensity map and the WMAP polarization analysis mask (PAM) as an example. The
PAM has an available sky fraction of $73\%$ (i.e. $f_{sky}=73\%$). Therefore
we construct P-filters with $P_t=35$ $\mu$K, whose $\mathbf{k} = 1$ regions
cover the same sky fraction as the PAM. We then let the rank $a=1, 2$, and
$10$ respectively to produce three different $\mathbf{k}$ maps for the
P-filter, and compare them with PAM in Figure~\ref{fig:p-filter1}, where PAM is
outlined by blue contour lines. The same procedure is applied to the Planck
353 GHz polarization intensity map together with the Planck 2018 component
separation mask (CSM), because CSM is mainly associated with 353 GHz. For
illustration, pixels that are 1 on both the $\mathbf{k}(a,P_t)$ map and
PAM/CSM are manually set to green.

As is evident from Figure~\ref{fig:p-filter1}, the regions where
$\mathbf{k}(a,P_t)=1$ have approximately the same morphology as the
corresponding WMAP/Planck masks, but there are also some different features
that are noteworthy. Firstly, outside PAM, one can see a few extended arches
that are not masked out by PAM, which does not change significantly with
different choices of $a$, while inside the mask the corresponding shape of the
filter critically depends on rank $a$. Variation of $a$ from 1 to 10 makes the
inner zone approach zero, in agreement with Figure~\ref{fig:p-filter}. 

Another feature of the P-filter is clearly visible along the Galactic plane at
the galactic longitudes $180\degree\le \textbf{l}\le 300\degree$ (right hand
side): For PAM, this zone is indiscriminately removed from analysis, while
with a P-filter, there is no strong suppression in this area
($\mathbf{k}\approx1$), because the polarization intensities here are not much
higher than the threshold $P_t$. Meanwhile, we can see many red dots outside
the PAM, which are not included in the PAM because, to reduce discontinuity,
it is not preferable to keep too many tiny holes in a top-hat mask. However,
for the P-filter family this is not a problem, because the filter function is
always continuous.

By comparing the upper and lower panels of Figure~\ref{fig:p-filter1}, we see
that, for the Galactic plane region at $180\degree\le \textbf{l}\le
300\degree$, the two top-hat masks (PAM and CSM) are similar, but the
corresponding P-filters are able to show more differences between 23 and 353
GHz: the P-filter based on 23 GHz gives less suppression in this region than
the one based on 353 GHz. Thus the P-filters automatically trace different map
signals and give more dedicated responses than top-hat masking. Also notice
that, in full sky, PAM and CSM are morphologically different, which means the
design of a common mask for multi-frequency analysis of the CMB foregrounds
and cosmological products need more investigations.

We also point out that the P-filter can be applied to both temperature and
polarization maps. An example for the temperature map can be found in
Appendix~\ref{app:P-filter for T-map}, which is similar to
Figure~\ref{fig:p-filter1}.

\section{Common P-filter for multiple frequency bands}\label{sec:how to make a common P-filter}

\subsection{Variation of the P-filter threshold}\label{sub:var of filter param vs freq}

\begin{figure}[!tbh]
  \centering
  \includegraphics[width=0.32\textwidth]{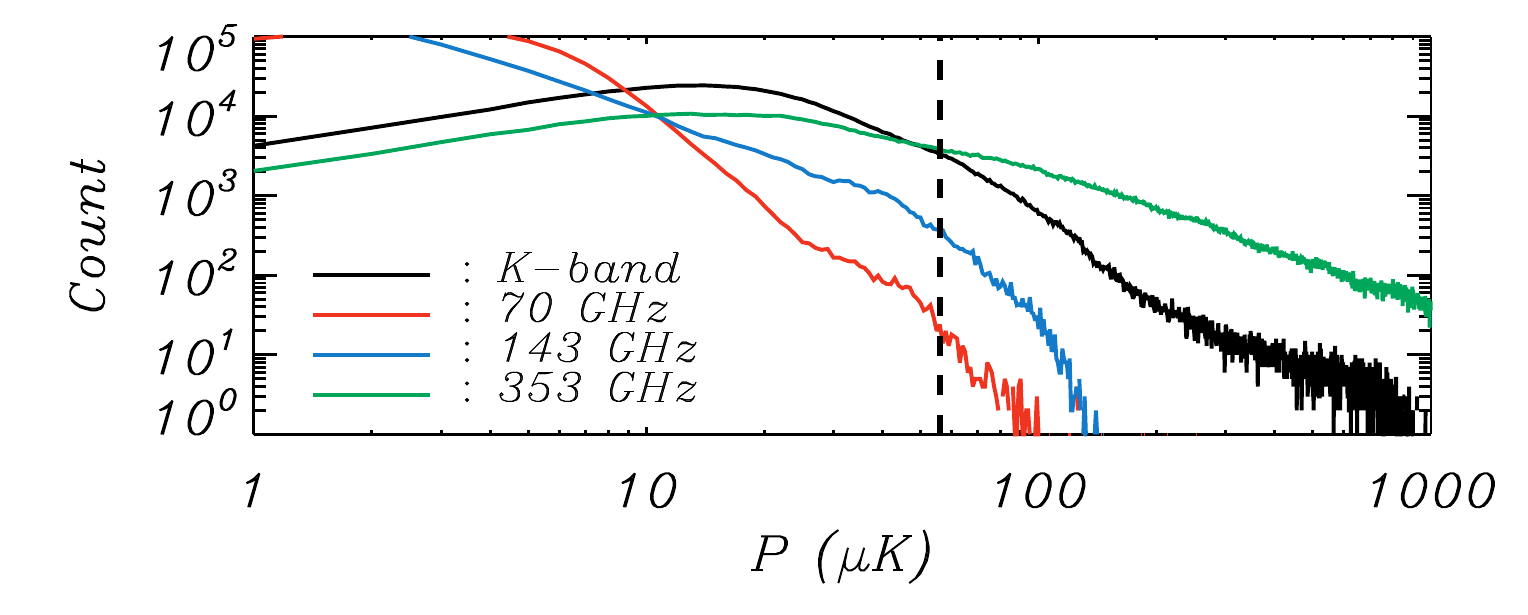}
  \includegraphics[width=0.32\textwidth]{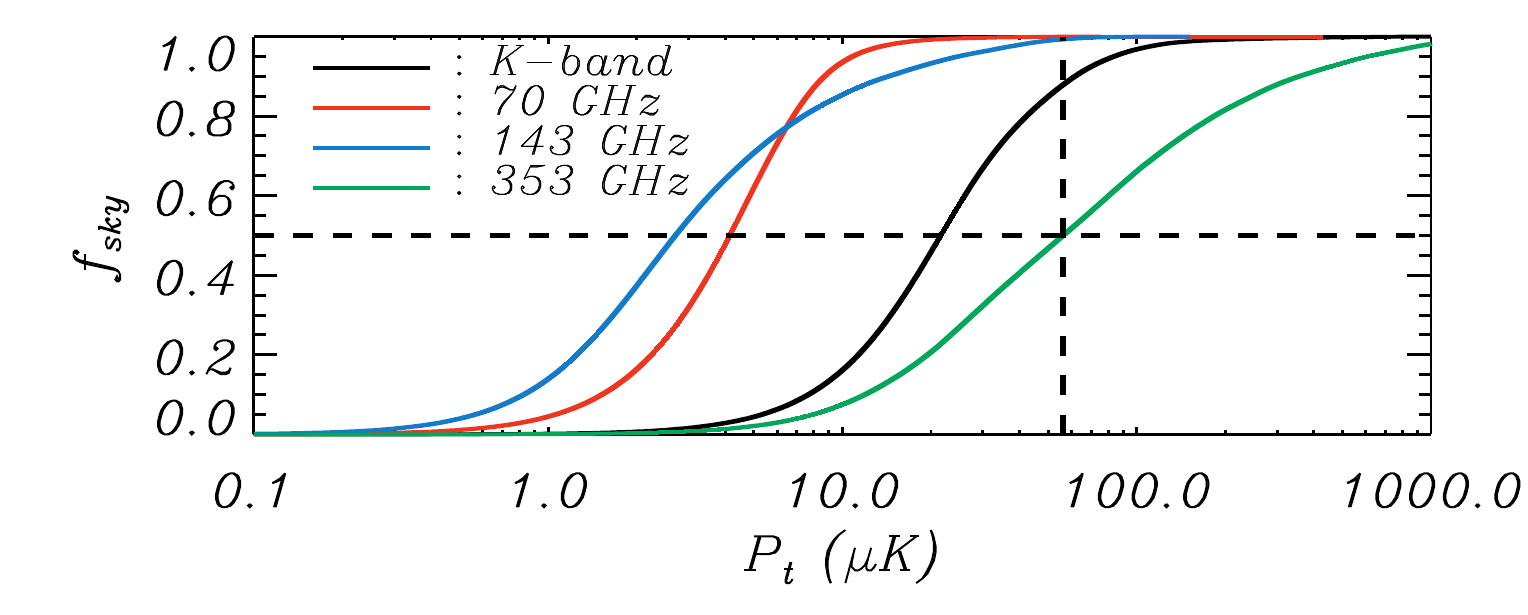}
  \includegraphics[width=0.32\textwidth]{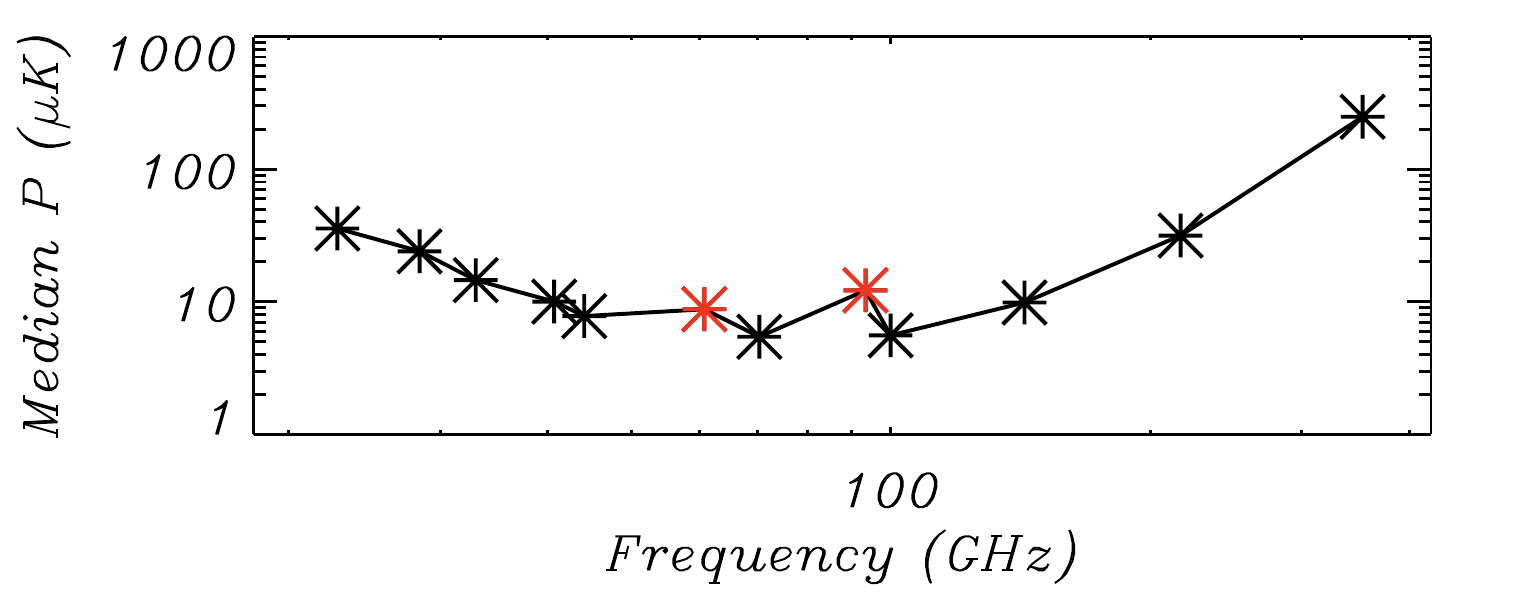}  
  \caption{ Illustrations of the relationship between $P_t$ and frequency
  bands. \emph{Left}: Histograms of the polarization intensities for some
  example frequency bands, where the vertical line marks 56 $\mu$K for
  example. \emph{Middle}: fraction of the sky ($f_{sky}$) with $P<P_t$ as
  function of $P_t$, the vertical line is the same as in the top panel, and
  the horizontal line marks the 50$\%$ position. \emph{Right}: The median
  polarization intensity within the Planck union mask VS. the frequency bands.
  The two red points are the WMAP V and W bands.}
  \label{fig:p-filter2}
\end{figure}

We argue here that, one cannot use the same P-filter threshold for two
different frequency bands, if they contain different foregrounds. To
illustrate this issue, we take under consideration the WMAP 23--94 and Planck
30--353 GHz bands (at $N_{side}=128$), and construct histograms of the
corresponding polarization intensity maps (before P-filtering). Some of these
histograms are shown in the top panel of Figure~\ref{fig:p-filter2}. They help
to illustrate how the threshold $P_t$ for each frequency band should be
determined differently. Taking the Planck 353 GHz band for example, in
Table~\ref{tbl:t1} we present the values of $P_t$ that gives the corresponding
sky fraction $f_{sky}$. The vertical line shows $P_t=56$ $\mu$K corresponding
to $f_{sky}=0.5$ from 353 GHz map. For K-band this threshold removes roughly
same amount of pixels, while for 70 and 143 GHz maps only a small fraction of
the pixels can exceed $P_t=56$ $\mu$K.

\begin{table}[!tbh]
 \caption{The sky fraction of $\mathbf{k}=1$ for the 353 GHz polarization
 intensity map with a given threshold $P_t$.}
 \centering
 \begin{tabular}{|l|r|r|r|r|} \hline
 $f_{sky}$, $\%$  & 50 & 60 & 70  & 80   \\ \hline
 $P_t$, $\mu$K    & 56 & 80 & 116 & 183  \\ \hline
 \end{tabular}
 \label{tbl:t1}
\end{table} 

The middle panel of Figure~\ref{fig:p-filter2} shows the dependency of the sky
fraction $f_{sky}$ on $P_t$ for the same bands in the left panel. From this
panel one can see that, for instance, keeping $f_{sky}\simeq 0.5$ requires
$P_t\simeq2-5$ $\mu$K for 70 and 143 GHz, and $P_t\simeq20$ $\mu$K for the K
band. The bottom panel of Figure~\ref{fig:p-filter2} illustrates the median
polarization intensities of the region covered by an example mask (Planck 2015
Union mask) for all the frequency bands, which also show the tendency of
polarization intensity variation in the given region versus frequencies.

\subsection{Construction of a common P-filter}\label{sub:construction of common P-filter}

We see from above that the P-filter should be different for different
frequency bands. However, it is still possible to generate a common P-filter
for all frequency bands.

Primarily, to determine the P-filter for all bands, one needs $2N_{band}$ free
parameters. However, we can reduce this number to \emph{only 1} as follows: We
first set $f_{sky}=0.8$ for all frequency bands as a possible variant (one is
free to choose another value based on any specific requirement). As we have
shown in the previous section, the fraction $f_{sky}$ will determine the
thresholds $P^{(i)}_t$, and with the corresponding ranks $ a^{(i)}_{max}$
determined by Eq.~\ref{equ:recomm rank}, one can fully determine the P-filter
for all bands with only 1 free parameter: $f_{sky}=0.8$. For convenience, we
list all $2N_{band}$ parameters generated from $f_{sky}=0.8$ in
Table~\ref{tab:trivial params}. One can see that for all sky maps the ranks
$a$ are close to 1.25, with some small variations like $a=1.33$ for 143 GHz
and 1.29 for 353 GHz.
\begin{table}[!htb]
 \caption{The  filter parameters used in Section~\ref{sub:construction of
 common P-filter}, which are all correspond to the sky fraction
 $f_{sky}=80\%$. The unit for $P_t$ is $\mu$K. }
 \centering
 \begin{tabular}{|c|c|c|c|c|c|c|c|c|c|c|c|c|} \hline
 Band    &      K &     30 &     Ka &      Q &     44 &      V\\ \hline
 $P_t$   &     42 &     20 &     17 &     13 &      9 &     12 \\ \hline
 $a$     &   1.21 &   1.22 &   1.21 &   1.22 &   1.22 &   1.27 \\ \hline \hline
 Band    &     70 &      W &    100 &    143 &    217 &    353 \\ \hline 
 $P_t$   &      6 &     16 &      4 &      7 &     23 &    180 \\ \hline
 $a$     &   1.24 &   1.38 &   1.26 &   1.33 &   1.29 &   1.29 \\ \hline
 \end{tabular}
 \label{tab:trivial params}
\end{table} 

\begin{figure}[!h]
  \centering
  \includegraphics[width=0.32\textwidth]{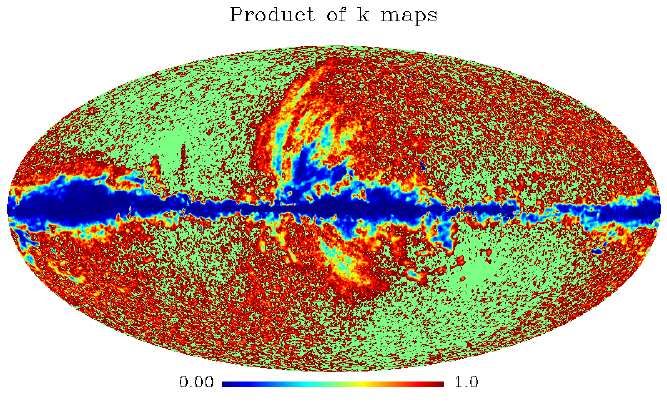}
  \includegraphics[width=0.32\textwidth]{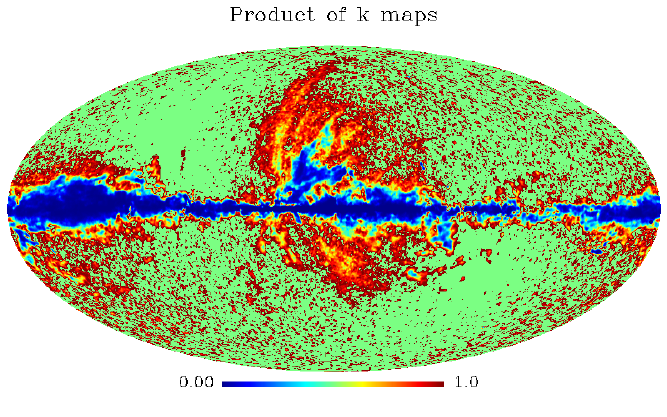}
  \includegraphics[width=0.32\textwidth]{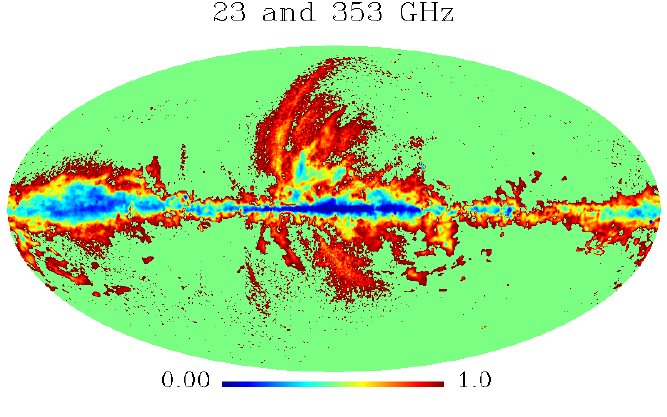}
  \caption{The $K(\textbf{n})$  maps produced from Table~\ref{tab:trivial
  params}. \emph{Left}: for all 12 bands. \emph{Middle}: for only the Planck
  frequency bands. \emph{Right:} for only the 23 and 353 GHz (synchrotron and
  dust), which is the one to be used in this work. The region where
  $k(\mathbf{n})=1$ is marked in green for visibility. }
  \label{fig:p-filter kmap all}
\end{figure}

Based on Table~\ref{tab:trivial params}, we can define a common filter
$\mathbf{K}$ as the multiplication of all 12 $\mathbf{k}_i$ (other
combinations are also possible) maps as
\begin{equation}\label{equ:trivial k map}
\mathbf{K} = \prod^N_{i=1} \mathbf{k}_i,
\end{equation}
which is shown in Figure~\ref{fig:p-filter kmap all}. Similarly, we can also
define the common filter using only the Planck bands or only the lowest and
highest bands (23 and 353 GHz), which are all shown in the same figure.

The combination of all 12 bands gives a very aggressive common P-filter,
which can be used for the most conservative purposes. A curious feature of the
$\mathbf{K}$-map that includes the WMAP bands is the presence of the
instrumental noise, clearly seen along the Ecliptic plane zone as red dots.
Nevertheless, the amplitudes of the $\mathbf{K}$-map in those dots are close
to 1 (which still ensures continuity). The $\mathbf{K}$-map for Planck bands
alone is less noisy, but it still reveals some features of the synchrotron and
thermal dust well outside the Galactic plane, associated with galactic Loop I
and arches~\citep{2015MNRAS.452..656V}. As it is seen from
Figure~\ref{fig:p-filter kmap all}, the effective common P-filter for Planck
data alone is significantly different in respect to the CSM. It is less
aggressive than CSM in the Galactic plane zone, but more aggressive in the
Loop I and surrounding zones. Recall that, the common P-filter is not fixed: it
can vary according to the combination of frequency bands included in
Eq.~\ref{equ:trivial k map}. Moreover, for the forthcoming CMB experiments the
particular shape of the common filter will be based on the new data, keeping
in mind, however, the recipe of its construction here and some peculiarities
of the filter to be listed below.

Below in Section~\ref{sec:what can pfilter do for us}, we will work with a
simplified case and use the $\mathbf{K}$-map produced only from the lowest and
highest bands (23 and 353 GHz), shown in the right panel of
Figure~\ref{fig:p-filter kmap all}. This can reflect the basic properties of
the synchrotron and dust polarizations and is least affected by the
instrumental noise in the polarized foregrounds.

\section{Advantages of a common P-filter}\label{sec:what can pfilter do for us}

As mentioned above, in this section we use the simplified common P-filter in
the right panel of Figure~\ref{fig:p-filter kmap all} for testing purposes.

\subsection{Automatic removal of the strong point sources}\label{sub:auto remove point source}

If the amplitude of a point source exceeds the threshold $P_t$ for some of the
frequency band maps included in the $K(\textbf{n})$ map, it will be
automatically removed by the P-filter for all the maps without any extra cost
and without losing continuity.

\begin{figure}[!h]
  \centering
  \includegraphics[width=0.32\textwidth]{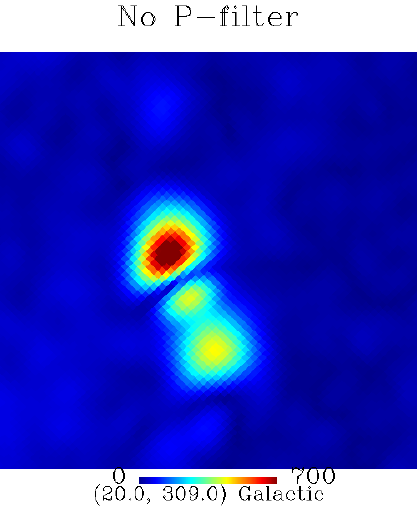}
  \includegraphics[width=0.32\textwidth]{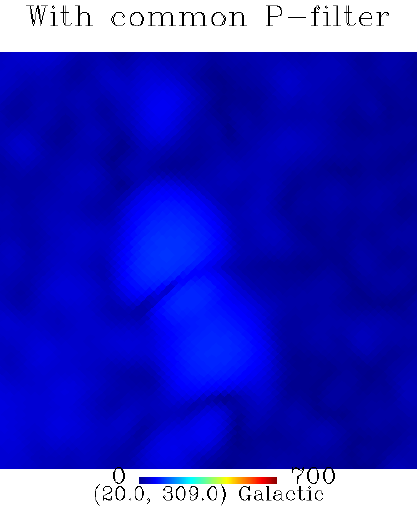}
  \caption{Example of how the common P-filter automatically and smoothly
  removes a polarized point source in the WMAP K-band map around
  $(\textbf{l},\textbf{b})=(309\degree,20\degree)$, without any special
  design. The map unit is $\mu$K (same for the rest sky maps).}
  \label{fig:auto remove ps}
\end{figure}

For example, with the common P-filter mentioned above (which makes no
additional requirements on the method of point source removal), the polarized
strong point sources in the WMAP K-band map around
$(\textbf{l},\textbf{b})=(309\degree,20\degree)$ are automatically removed, as
shown in Figure~\ref{fig:auto remove ps}. Note that the removal will still
maintain continuity of the signal in the map, and obviously a better and more
thorough point source removal can be done by adjusting the filter parameters
for a specific band and a specific local region.

\subsection{Reducing the leakage associated with bandpass}\label{sub:example with bandpass}

In Fourier analysis, it is well known that when a localized signal is cut off
in the frequency domain, it will no longer be localized anymore, but instead will
start to leak into other regions in the pixel domain, even those far away from the source.
In the analysis of a sky map, a similar problem also exists: when a sky map is
bandpassed in the harmonic space, the high amplitude signal from the
brightest pixels will start to leak into other regions and produce unwanted
contamination.

There are several ways to alleviate this kind of leakage due to bandpassing,
from simply using a top-hat mask, to harmonic space apodization (see for
example~\cite{2016A&A...594A...9P}) or wavelets (see for
example~\citep{2012MNRAS.419.1163B}). However, these techniques can exhibit
some problems. For example, masking is the simplest solution, but it will
generate further Gibbs-like effects, and it will cause other problems like
power spectrum leakage~\citep{2002ApJ...567....2H} and
EB-leakage~\citep{2006PhRvD..74h3002S, 2010A&A...519A.104K,
PhysRevD.82.023001, 2017PhRvD..96d3523B, 2018arXiv180105358K}. Harmonic space
apodization will inevitably keep some unwanted multipole components;
meanwhile, there are countless schemes for apodization, which makes it
difficult to make a ``natural'' choice. Wavelets are computationally
expensive, and they do not have straightforward mapping to a single spherical
harmonic component.

The P-filter provides an easy, quick and cheap way to prevent the leakages
right from the pixel domain. To provide an example, in Figure~\ref{fig:p-filter
maps}, we show the effect of the above mentioned common P-filter in
combination with a simple bandpass from $L=60$ to $L=100$ on the 353 GHz
polarization map (P-filtering first, then bandpassing). One can see that,
without the P-filter, the bandpass will cause serious stripe-like leakages,
but after applying the common P-filter introduced in the end of
Section~\ref{sub:construction of common P-filter}, the leakage is largely
eliminated. This example can be further improved by optimizing the common
P-filter based on more specific requirements.
\begin{figure}[!htb]
  \centering
  \includegraphics[width=0.45\textwidth]{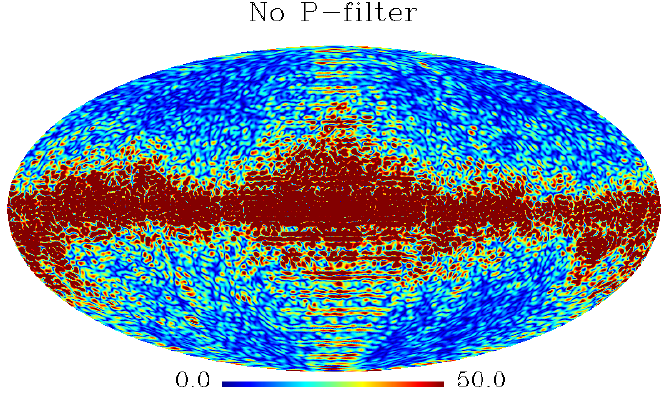}
  \includegraphics[width=0.45\textwidth]{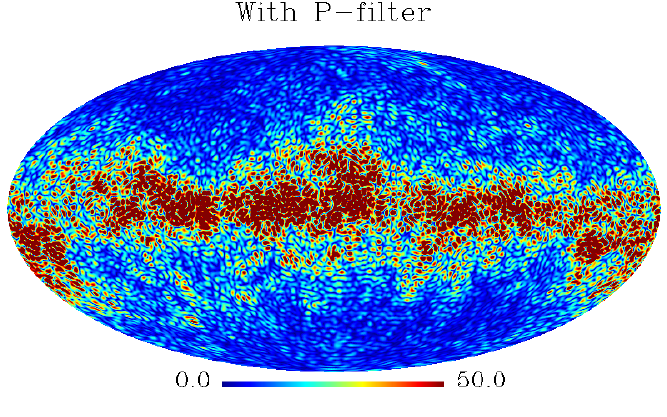}

  \includegraphics[width=0.45\textwidth]{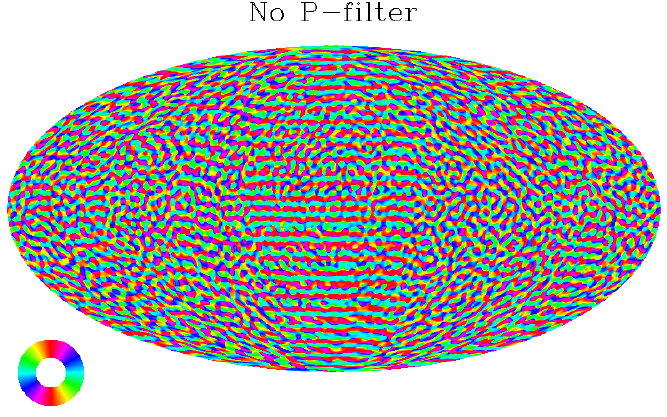}
  \includegraphics[width=0.45\textwidth]{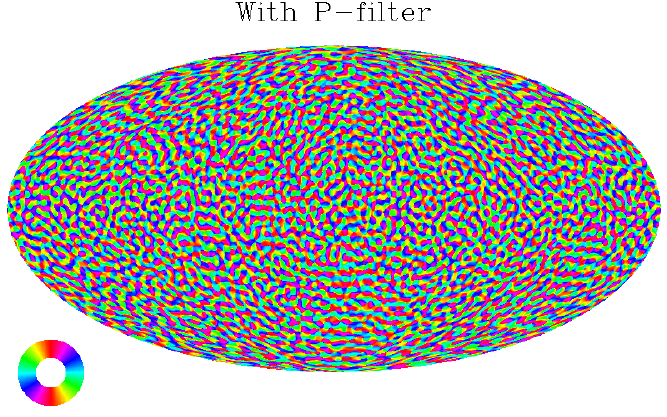}
  \caption{Results of bandpassing the Planck 353 GHz polarization map from
  $L=60$ to $L=100$ without/with (left/right) the common P-filter in advance.
  \emph{Upper}: polarization intensities. \emph{Lower}: polarization angles.}
  \label{fig:p-filter maps}
\end{figure}

\subsection{Reducing the EE, BB leakages}\label{sub:help improve EE and BB leakage}

A full sky polarization map can be decomposed
into $(Q_E,U_E)$ and $(Q_B,U_B)$ families of Stokes parameters that come only
from the E or B modes respectively. This decomposition satisfies
\begin{eqnarray}\label{equ:eb family}
(Q,U)\equiv(Q_E,U_E)+(Q_B,U_B),
\end{eqnarray}
where $(Q,U)$ are the input Stokes parameter maps. More details can be found
in Appendix~\ref{app:eb family} and~\cite{2018JCAP...05..059L,
2018A&A...617A..90L}.

\begin{figure}[!htb]
  \centering
  \includegraphics[width=0.45\textwidth]{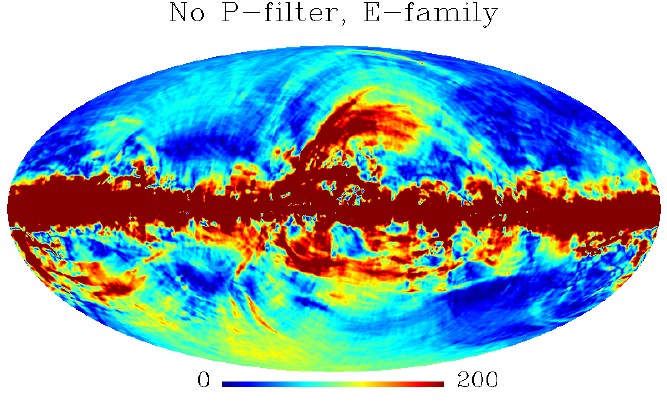}
  \includegraphics[width=0.45\textwidth]{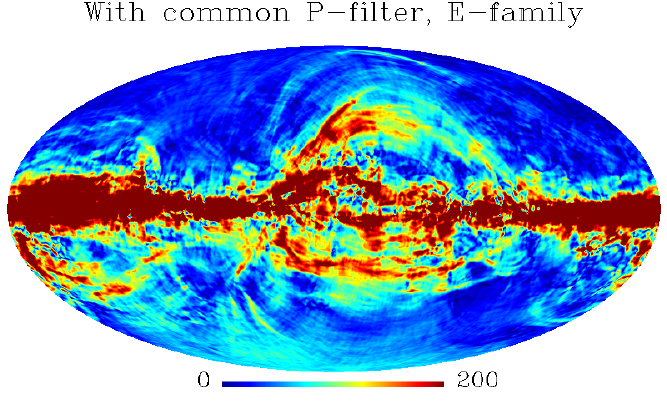}

  \includegraphics[width=0.45\textwidth]{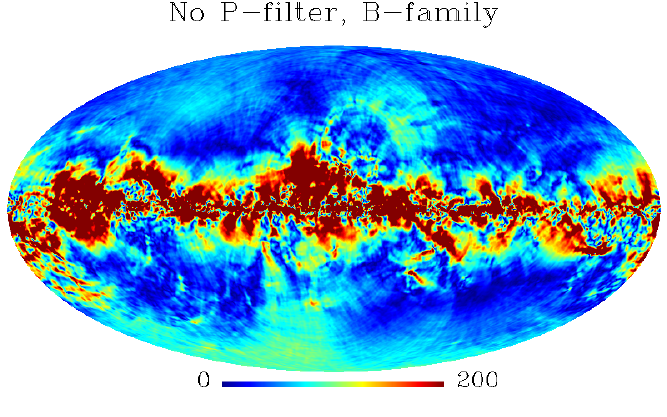}
  \includegraphics[width=0.45\textwidth]{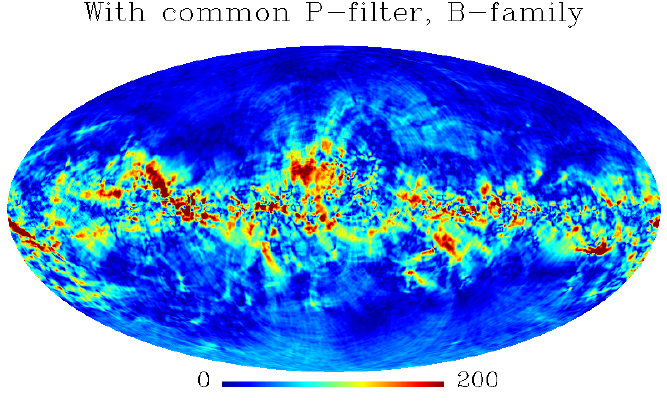}
  \caption{Full sky E/B family components (upper/lower) derived from the
  Planck 353 GHz map before/after (left/right) applying the common P-filter.}
  \label{fig:p-filter map no bandpass}
\end{figure}

If a polarized sky map is decomposed into E and B families (or traditional E
and B modes) without a P-filter, then there is going to be cross-talk between
different regions of the sky, even if one uses a full sky map from the
beginning. The reason is that, very bright regions (like the Galactic plane)
will produce many extended pixel domain $(Q_E,U_E)$ and $(Q_B,U_B)$ family
components, which are pure contamination for other regions. Note that this
phenomenon is unique for polarization, and does not exist for a temperature
map, except for some small residuals due to the pixelization and $L_{max}$
limitation. One important advantage of the P-filter is to reduce such EE and
BB leakages in estimating the full sky $(Q_E,U_E)$ and $(Q_B,U_B)$ families.
The reason is easy to understand: the brightest pixels will be significantly
suppressed by the P-filter in a \emph{continuous} way,

In Figure~\ref{fig:p-filter map no bandpass}, we present straightforward
comparisons of decomposing the full-sky Planck 353 GHz polarization map into
the $(Q_E,U_E)$ and $(Q_B,U_B)$ families with and without the common P-filter
mentioned above. The prevention of leakage can be easy observed by much
cleaner results at higher latitudes.

\subsection{Reducing the EB-leakage}\label{sub:eb leakage}

Formally speaking, calculation and separation of the E and B families
(Eq.~\ref{equ:eb family}) should be done on the full sky. If part of the sky
is missing, the two families will be mixed and create E-to-B and B-to-E
leakage in the available sky region. For CMB science, one is mainly interested
in the E-to-B leakage, because the B-mode of CMB is related to the primordial
gravitational waves.

If, from the very beginning, only part of the sky is available, then the
E-to-B leakage cannot be precisely estimated. However, we can construct tests
using full-sky foreground maps as known inputs to simulate and evaluate the
E-to-B leakage. Here we perform such a test using the P-filter.

\begin{figure}[!tbh]
  \centering
  \includegraphics[width=0.45\textwidth]{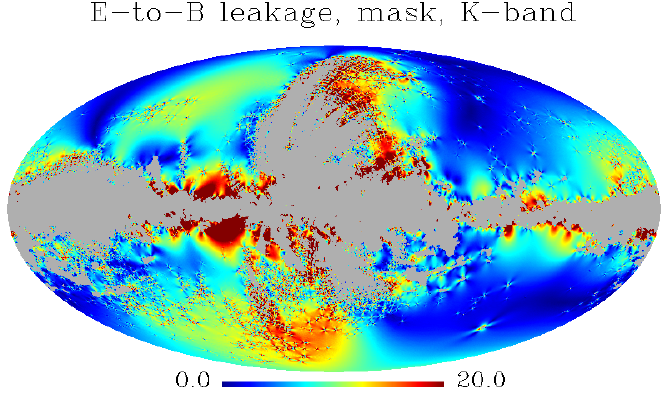}
  \includegraphics[width=0.45\textwidth]{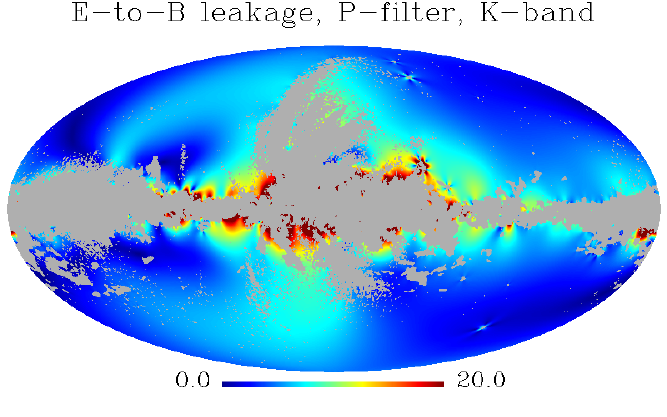}

  \includegraphics[width=0.45\textwidth]{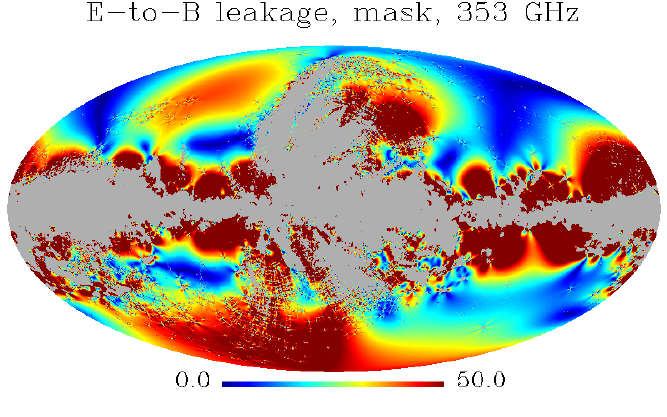}
  \includegraphics[width=0.45\textwidth]{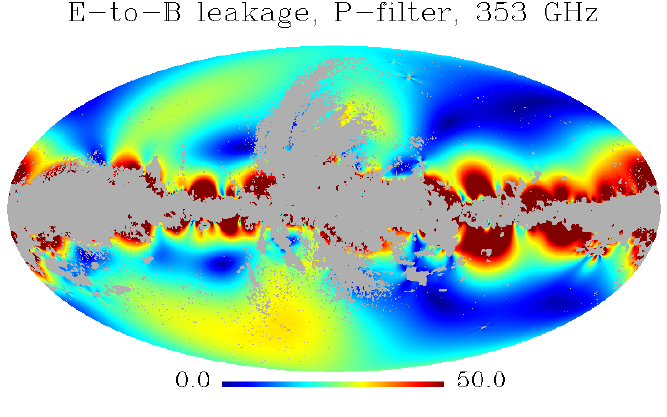}
  \caption{Examples showing how a common P-filter helps to alleviate the
  E-to-B leakage (without further correction). \emph{Upper}: K-band.
  \emph{Lower}: 353 GHz. \emph{Left}: with a top-hat mask. \emph{Right}: with
  a P-filter. }
  \label{fig:p-filter eb1234}
\end{figure}

We take the WMAP K-band full-sky synchrotron polarization maps and separate
them into E- and B-families as described in Appendix~\ref{app:eb family}. Then
we apply a top-hat mask to the E-family and redo the EB-separation. The
difference between the resulting B-family in the available sky region and the
known real B-family is exactly the E-to-B leakage due to the top-hat masking.
We then repeat the procedure using the common P-filter instead of a top-hat
mask and compare the results in Figure~\ref{fig:p-filter eb1234}. We can see
that, without any correction, the P-filter gives much lower E-to-B leakage
than the top-hat mask. Similar tests are done for the Planck 353 GHz map,
giving similar results as shown in the same figure. Therefore, although a
top-hat mask completely kills all foreground in the ``unwanted'' region, a
P-filter with moderate suppression will actually give much lower E-to-B
leakage and a better B-mode in the ``desired'' region. We use a top-hat mask
as a reference here and below to show that the P-filter does not suffer from
the same problems related to the top-hat masking. Comparisons to more
realistic masking techniques used in CMB analysis, such as masks with smoothed
boundaries, will be given in a future paper.

\subsection{Reducing the EB-leakage for a CMB map}\label{sub:with bandpass cmb}

When the P-filter is applied to a band map, not only the foregrounds, but also
the underlying CMB signal will be multiplied by the common $\mathbf{K}$-map.
This could potentially change the polarization angles of the $(Q_E,U_E)$ and
$(Q_B,U_B)$ families, which are defined as
\begin{align}
\theta_E &= 0.5\mathbf{arctan2}(U_E,Q_E), \\ \nonumber
\theta_B &= 0.5\mathbf{arctan2}(U_B,Q_B),
\end{align}
where $\mathbf{arctan2}$ is an extension of the $\mathbf{arctan}$ function
that returns results in the range $(0,2\pi)$. Here we check the extent to
which $\theta_E$ and $\theta_B$ might change due to the P-filter. We generate
a simulated CMB map, filter it with the common P-filter, and bandpass to
$\ell=60$--100. Then we extract the polarization angles $\theta_E'$ and
$\theta_B'$ in interior of the region with $\mathbf{K}=1$, where we discard
the edges by excluding a 3-pixel width contour (at $N_{side}=512$) along the
edge of the mask. Then we compute the following quantities to compare the
angles of the filtered map with those of the real CMB, (denoted by $\theta_E$
and $\theta_B$):
\begin{align}
\Delta\theta_E&=|\arcsin(\theta'_E-\theta_{E})|, \\ \nonumber
\Delta\theta_B&=|\arcsin(\theta'_B-\theta_{B})|,
\end{align}
where $\theta'_E$ and $\theta'_B$ are the polarization angles after filtering.
Note that because the change $\Delta\theta_B$ is dominated by the E-to-B
leakage, this test shows the extent to which the P-filter helps to alleviate
the E-to-B leakage problem.

\begin{figure}[!t]
  \centering
  \includegraphics[width=0.96\textwidth]{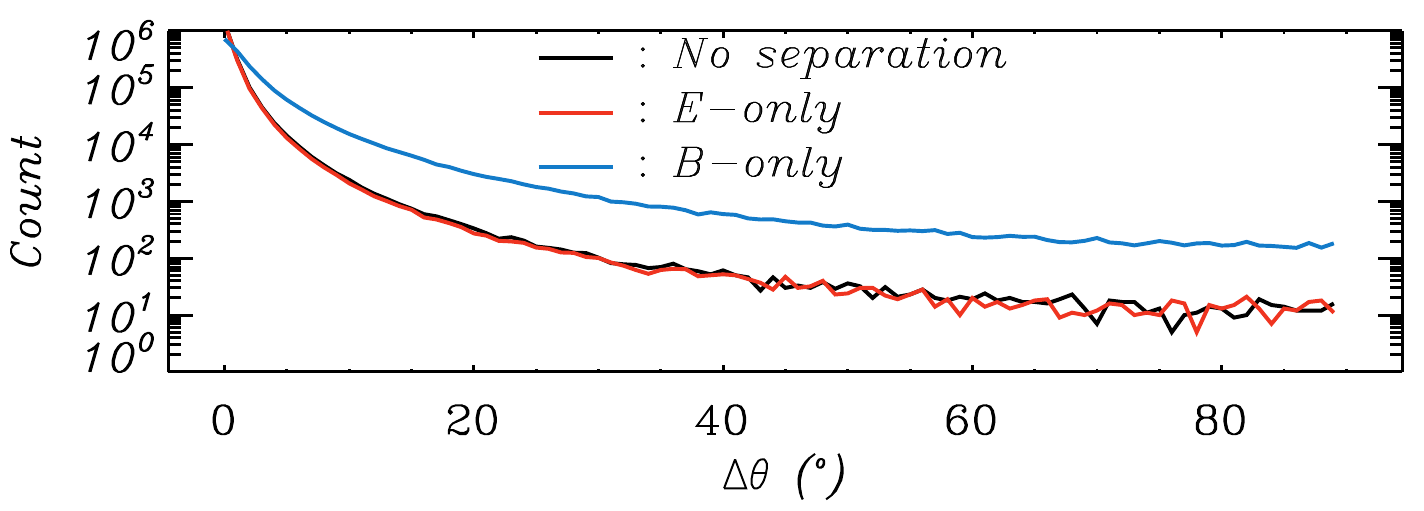}
  \caption{Histogram of the polarization angle differences after
  the pixel domain filtering using the common P-filter and band passing for
  $L=60$--$100$, for no separation, $\Delta\theta_E$ and $\Delta\theta_B$
  respectively. The mean polarization angle variations are $1.0\degree
  (5.5\degree)$ for no separation and $\Delta\theta_E$, and is $3.0\degree
  (17\degree)$ for $\Delta\theta_B$, where the values in brackets are for the
  reference case with a top-hat mask.}
  \label{fig:p-filter hist pol ang diff}
\end{figure}

We calculate the histogram of the changes of $\Delta\theta_E$ and
$\Delta\theta_B$ and plot them in Figure~\ref{fig:p-filter hist pol ang diff}.
We also calculate the mean polarization angle changes, which are only
$1.0\degree$ for no separation and $\Delta\theta_E$, and only $3.0\degree$ for
$\Delta\theta_B$, compared to $5.5\degree$ and $17\degree$ respectively for
the case with a top-hat mask. Therefore, with a P-filter, the variation of the
polarization angle is significantly reduced in comparison to using a top-hat
mask.

\section{Example of optimizing the filter parameter}\label{sec:EB-leakage and optimization}

Here we discuss a more practical situation, where we shall apply the P-filter
on a CMB+foreground map directly (without foreground removal), and try to
optimize the filter to suppress the leakage of foreground and obtain a CMB
B-mode signal that is as reliable as possible in the sky region of interest.
Especially, since the P-filters are smooth and continuous, we expect it to
give much less E-to-B leakage in the region of interest. Note that for faster
optimization, we use $N_{side}=128$ in this section. Also note that here we
allow the filter parameters to be variables, so we are not dealing with the
common P-filter in Section~\ref{sec:how to make a common P-filter} anymore.
Finally, since we are studying the P-filter in this work, we do not use any
other correction methods.

\begin{figure}[!htb]
  \centering
  \includegraphics[width=0.32\textwidth]{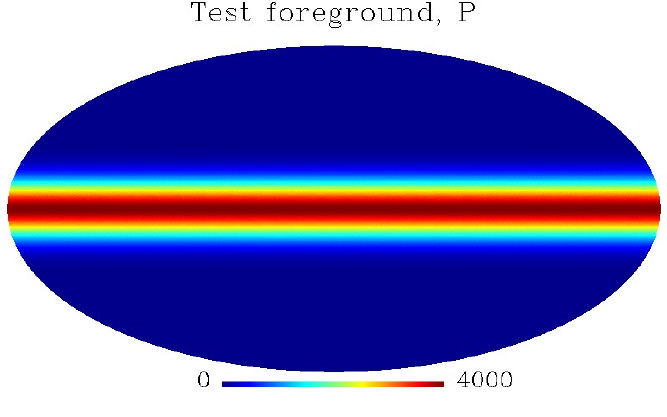}
  \includegraphics[width=0.32\textwidth]{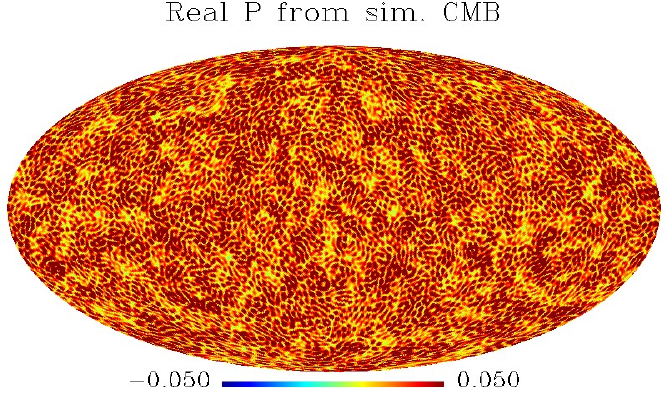}
  \includegraphics[width=0.32\textwidth]{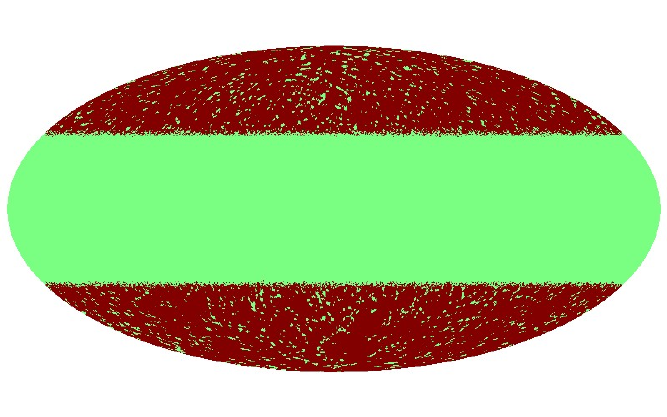}

  \includegraphics[width=0.48\textwidth]{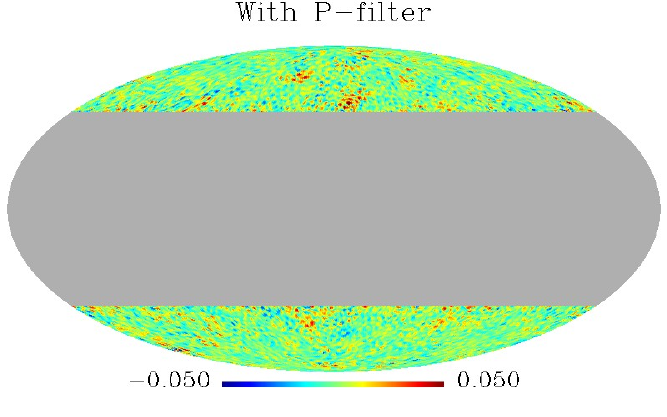}
  \includegraphics[width=0.48\textwidth]{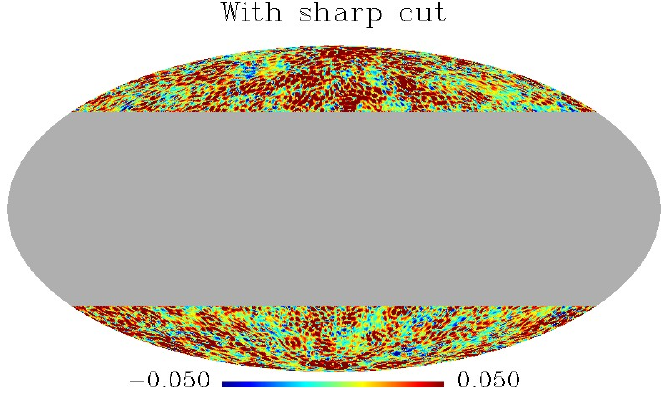}
  \caption{\emph{Upper, from left to right}: polarization intensity of the
  input test foreground, real $\mathbf{P}_B$ map from the input CMB map in
  multipole range $L=60$--$100$, and illustration of the common working region
  of the P-filter (green). \emph{Lower left}: the map of $P_B$ difference,
  computed as $\mathbf{P}_B-\mathbf{P}_B^{real}$ in the multipole range
  $L=60$--$100$. \emph{Lower right}: same as the lower-left panel, only that
  top-hat masking is used instead of the P-filter. The ``region of interest''
  is the non-gray region in the lower panels. See Section~\ref{sec:EB-leakage
  and optimization} for more details.}
  \label{fig:p-filter cmb and fg}
\end{figure}

A detailed description of the processes is as follows:
\begin{enumerate}
\item We generate a test foreground that has a belt shape along the Galactic
plane, which decays exponentially at higher latitudes with FWHM = $20\degree$.
The central polarization intensity is 4000 $\mu$K, and the polarization angles
are randomly and uniformly distributed. The test foreground is shown in the
upper-left panel of Figure~\ref{fig:p-filter cmb and fg}.

\item We generate a simulated CMB map with $r=0.05$, and use the
CMB+foreground map as the input of the P-filter. Note that the real CMB map
will not be used anymore until the final comparison.

\item We take a ``region of interest'' that has $|\textbf{b}|>45\degree$,
to simulate using high Galactic latitude region. This ``region of interest''
is only used for comparison of the final results, and one can see later from
Figure~\ref{fig:p-filter cmb and fg} that the comparison is not sensitive to the
size of this region.

\item \label{itm:optimal param} We try several P-filters with various
threshold $P_t$ and rank $a$ on the CMB+foreground map, and for each set of
$(P_t,a)$, we calculate the filtered B-mode polarization intensity
($\mathbf{P}_B=\sqrt{\mathbf{Q}_B^2+\mathbf{U}_B^2}$) from the full sky, and
then calculate the average polarization intensity
$\<P_B\>=\sqrt{\<\mathbf{P}_B^2\>}$ only in the ``region of interest''. The
best threshold and rank are determined by minimizing $\<P_B\>$. The results
are $P_t=1.15$ $\mu$K and $a=3$. Note that this step does not require to know
the input CMB or foreground, and the optial parameters are specific to the
simulated foreground in use.

\item \label{itm:compare1} With $P_t$=$1.15$ $\mu$K and $a=3$, we filter the
CMB+foreground map, and derive $P_B$ from the filtered full sky map in the
multipole range from $L=60$ to $L=100$. Then we subtract the known CMB $P_B$
from the filtered $P_B$, and show the difference in the lower-left panel of
Figure~\ref{fig:p-filter cmb and fg}. Note that the region with $\mathbf{k}\ne1$
is the region in which the P-filter really works, which is called the
``working region''. The ``working region'' is also shown in the upper right
panel of the same figure.

\item We use the same threshold $P_t$ but infinite rank $a$ to repeat
step~\ref{itm:compare1}, which represents the performance of a top-hat mask.
The corresponding $\mathbf{P}_B$ map difference is presented in the
lower-right panel of Figure~\ref{fig:p-filter cmb and fg}.
\end{enumerate}

Except for the final comparison of the $\mathbf{P}_B$ map, the above procedure
depends only on the combined CMB+foreground map. From Figure~\ref{fig:p-filter
cmb and fg}, we can see that, a P-filter can significantly suppress the
leakage due to the foreground and output a B-mode polarization intensity very
similar to the real input in the ``region of interest''. On the other hand, a
top-hat mask certainly kills the foreground more thoroughly, but the price is
to leave much more leakage in the whole ``region of interest'', which is not
preferable.

\section{Discussion}\label{sec:discussion}

In this work, we design the P-filter, which succeeds the basic idea of masking
in CMB science to ``remove the brightest pixels''. Instead of top-hat masking,
the P-filter works in a smooth and continuous way, with simple and clear
mathematical definition.

The P-filters help to greatly simply the task to design a mask for analysis:
For a top hat mask, one has to pay extra attention to the serrated edge and
the individual tiny holes, and, to fix them, one often need some arbitrary
choice. However, for a P-filter, the serrated edge and the individual tiny
holes are not a problem at all, and the filter can be designed with just one
parameter\footnote{If one choose $a_{max}$ given by Eq.~\ref{equ:recomm
rank}}. It is also much easier to use P-filter than to use a top-hat mask when
working with multiple resolutions, because for a top-hat mask, changing its
resolution is often problematic, but a P-filter can be conveniently
regenerated at a new resolution using the same set of parameter(s).

The pixel domain form of the P-filter is same as a pixel domain window
function/apodization, but they have completely different idea of design. A
normal pixel domain window function is defined on independent variables like
sky coordinates $(\theta,\phi)$, or time $t$ (for a time series). However, the
P-filter is defined directly on dependent variable like polarization intensity
or temperature. This is why the P-filter is independent to the coordinate
system, and is able to trace the map morphologies automatically.

The idea to work on dependent variables also brings two other advantages on
continuity and gradient. Firstly, we know that real maps have to be presented
with a finite resolution, thus all window functions defined on the coordinate
system (independent variables) will be affected and will bring extra
discontinuity depending on the way they are defined (although this effect is
normally small). However, the P-filter is defined on the dependent variables,
thus it will not bring extra discontinuity due to pixelization. Secondly,
window functions defined on independent variables can either increase or
decrease the gradient, however, a P-filter will always help suppress the high
gradients due to strong foregrounds, which is convenient.

We show in this work that the P-filter family can effectively reduce many
leakage problems, including the ones associated with bandpassing and
EB-separation. It can also be used to naturally suppress the regions with
residual contamination in CMB products or a partial sky map, for example,
removal of bright point sources. However, we point out that the P-filter can
not resolve the problem of peculiar zones of the maps, especially the ones
related to the systematic effects, which should always be treated separately
for safety.

The P-filter family has a well-defined semi-analytic form. As shown in
Section~\ref{sec:how to make a common P-filter}, when applied to multiple
frequency bands to create a common P-filter, it is possible to reduce the
number of free parameters to only 1, which is the $f_{sky}$ with
$\mathbf{k}(a,P_t)=1$. Note that this is exactly the parameter that represents
the basic idea of masking in CMB science.

The family of P-filters also provides a mathematical means to move smoothly
from the case of no filtering to the case of top-hap masking, by continuously
changing the rank parameter $a$ from 0 to $\infty$, as illustrated in
Figure~\ref{fig:p-filter}. For each intermediate step in the movement,
continuity of the filter is always preserved to the first derivative. This
provides an excellent way to find a balance between removing more contaminated
pixels and keeping more continuity by optimizing the threshold $P_t$ and the
rank $a$, as shown by the example in Section~\ref{sec:EB-leakage and
optimization}. Mathematically, this also provides a progressive way to study
the properties of top-hat masking by letting $a$ approach infinite, which
might be interesting.

\Ack{

This research has made use of data/product from the
WMAP~\cite{WMAPdata:online} and Planck~\citep{Planckdata:online}
collaboration. Some of the results in this paper are derived using the
HEALPix~\citep{2005ApJ...622..759G} package. This work was partially funded by
the Danish National Research Foundation (DNRF) through establishment of the
Discovery Center and the Villum Fonden through the Deep Space project. Hao Liu
is also supported by the National Natural Science Foundation of China (Grants
No. 11653002, 1165100008), the Strategic Priority Research Program of the CAS
(Grant No. XDB23020000) and the Youth Innovation Promotion Association, CAS.

}

\appendix

\section{The definition of EB-families}\label{app:eb family}

In our work, the E and B families refers to $(Q_E,U_E)$ and $(Q_B,U_B)$ that
contain only E or B mode respectively, and satisfy
$(Q,U)\equiv(Q_E,U_E)+(Q_B,U_B)$. They are defined as follows:
\begin{align}
\begin{split}\label{equ:EB-QU_final simple}
\begin{pmatrix}
Q_E \\
U_E
\end{pmatrix}(\bm{n})&= \int
\begin{pmatrix}
G_1  & +G_2 \\
+G_3  & G_4
\end{pmatrix}
(\bm{ n},\bm{ n}')
\begin{pmatrix}
Q \\
U
\end{pmatrix}(\bm{n}')\, d \bm { n}' \\
\begin{pmatrix}
Q_B \\
U_B
\end{pmatrix}(\bm{ n})&= \int
\begin{pmatrix}
G_4  & -G_3 \\
-G_2  & G_1
\end{pmatrix}
(\bm{ n},\bm{ n}')
\begin{pmatrix}
Q \\
U
\end{pmatrix}(\bm{ n}')\,d \bm {n}',
\end{split}
\end{align}
where the $G_{1-4}$ functions are defined as:
\begin{align}
\begin{split}G_{1}(\bm{ n},\bm{ n}') &= \sum_{l,m} F_{+,\ell m}(\bm{
    n})F^*_{+,\ell m}(\bm{ n}'), \\
    \quad G_{2}(\bm{ n},\bm{ n}') &= \sum_{l,m} F_{+,\ell m}(\bm{
    n})F^*_{-,\ell m}(\bm{ n}'), \\
    G_{3}(\bm{ n},\bm{ n}') &= \sum_{l,m} F_{-,\ell m}(\bm{
    n})F^*_{+,\ell m}(\bm{ n}'),\\ \quad G_{4}(\bm{ n},\bm{ n}') &=
    \sum_{l,m} F_{-,\ell m}(\bm{ n})F^*_{-,\ell m}(\bm{ n}'),
    \end{split}
\end{align}
and the $F_{+,-}$ functions are defined in terms of the spin-2 spherical harmonics
as:
\begin{align}
\begin{split}
\label{equ:define F}
    F_{+,\ell m}(\bm{ n}) &= -\frac{1}{2} \left[{}_{2}Y_{\ell m}(\bm{ n}) +
    {}_{-2}Y_{\ell m}(\bm{ n})
    \right], \\
    F_{-,\ell m}(\bm{ n}) &= -\frac{1}{2i}
    \left[{}_{2}Y_{\ell m}(\bm{ n}) - {}_{-2}Y_{\ell m}(\bm{ n}) \right].
\end{split}
\end{align}
Note that $G_i$ are real and $G_2 = G_3$.
For more details of these two families, see \cite{2018JCAP...05..059L,
2018A&A...617A..90L}.

\section{Example of P-filter for a temperature map}\label{app:P-filter for T-map}

\begin{figure}[!tbh]
  \centering
  \includegraphics[width=0.48\textwidth]{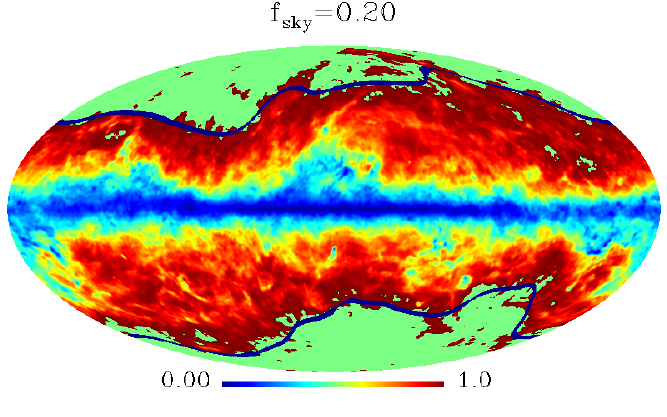}
  \includegraphics[width=0.48\textwidth]{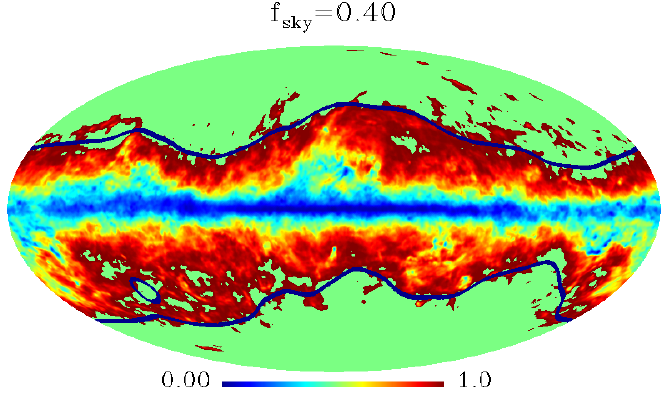}

  \includegraphics[width=0.48\textwidth]{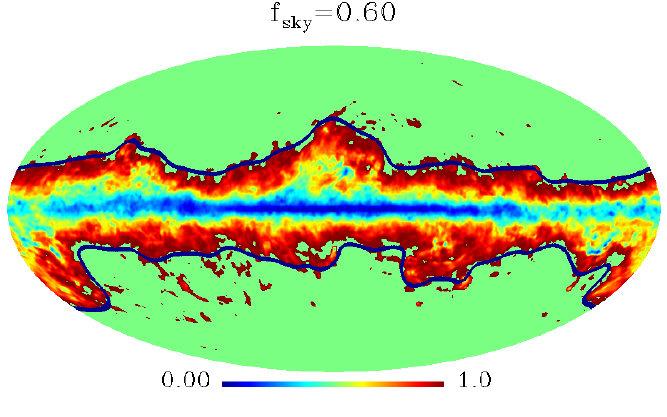}
  \includegraphics[width=0.48\textwidth]{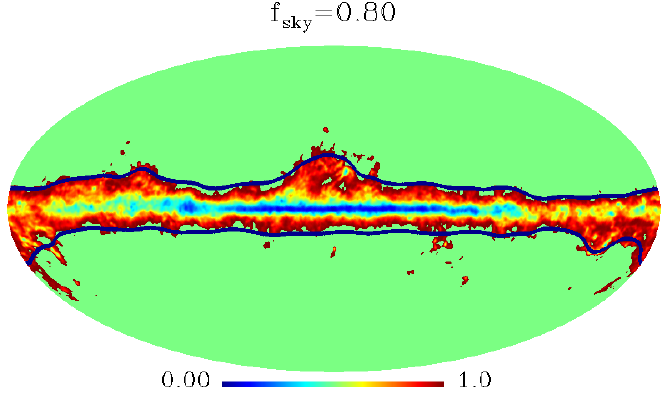}
  \caption{The $\mathbf{k}$ maps generated from the Planck 353 GHz
  temperature maps with $f_{sky}=0.2, 0.4, 0.6$, and $0.8$ (upper left to
  lower right), in comparison to the Planck power spectrum analysis masks with
  the same $f_{sky}$ (blue contour lines).}
  \label{fig:p-filter other masks}
\end{figure}

In Figure~\ref{fig:p-filter1}, we already see examples that the P-filter can
naturally provide the same sky coverage and similar morphology to existing
WMAP and Planck polarization masks. We also provide examples of generating
P-filters based on the Planck 353 GHz temperature (not polarization) maps, and
compare the results to the Planck power spectrum analysis masks\footnote{File
name: HFI\_Mask\_GalPlane-apo0\_2048\_R2.00.fits. From the file name, one can
see that these masks are based on the Planck HFI channels.}. For those masks,
Planck provides 8 different sky coverage with $f_{sky}=$0.2, 0.4, 0.6, 0.7,
0.8, 0.9, 0.97, and 0.99 respectively. For convenience, we take only
$f_{sky}=$0.2, 0.4, 0.6, and 0.8 to generate the corresponding variants of the
P-filters, and the polarization intensity P is replaced by the
temperature\footnote{The estimated CMB signal is removed in advance, so $T$ is
almost always positive.}, $T$, and the threshold $P_t$ by $T_t$. In this
paragraph, we focus on the effect the threshold has, thus we keep using the
same rank $a=1$. In Figure~\ref{fig:p-filter other masks} we compare the derived
$\mathbf{k}$ maps to the Planck masks in a similar way to
Figure~\ref{fig:p-filter1}, where one can see that for either case (30 or 353
GHz), the generated $\mathbf{k}$ maps have similar morphologies to the Planck
masks with the same $f_{sky}$. This tells us that the P-filters can easily
serve as an alternative of the masks used by WMAP and Planck to suppress the
foreground or residual foreground, and at the same time, automatically provide
apodizations that are continuous to the first derivative \emph{everywhere}.

\begin{thebibliography}{10}

\bibitem{2014A&A...571A...1P}
{Planck Collaboration}, P.~A.~R. {Ade}, N.~{Aghanim}, M.~I.~R. {Alves},
  C.~{Armitage-Caplan}, M.~{Arnaud} et~al., \emph{{Planck 2013 results. I.
  Overview of products and scientific results}},
  \href{https://doi.org/10.1051/0004-6361/201321529}{\emph{\aap} {\bfseries
  571} (Nov., 2014) A1}, [\href{https://arxiv.org/abs/1303.5062}{{\ttfamily
  1303.5062}}].

\bibitem{2016A&A...594A...1P}
{Planck Collaboration}, R.~{Adam}, P.~A.~R. {Ade}, N.~{Aghanim}, Y.~{Akrami},
  M.~I.~R. {Alves} et~al., \emph{{Planck 2015 results. I. Overview of products
  and scientific results}},
  \href{https://doi.org/10.1051/0004-6361/201527101}{\emph{\aap} {\bfseries
  594} (Sept., 2016) A1}, [\href{https://arxiv.org/abs/1502.01582}{{\ttfamily
  1502.01582}}].

\bibitem{2018arXiv180706205P}
{Planck Collaboration}, Y.~{Akrami}, F.~{Arroja}, M.~{Ashdown}, J.~{Aumont},
  C.~{Baccigalupi} et~al., \emph{{Planck 2018 results. I. Overview and the
  cosmological legacy of Planck}}, {\emph{ArXiv e-prints} (July, 2018) },
  [\href{https://arxiv.org/abs/1807.06205}{{\ttfamily 1807.06205}}].

\bibitem{2012SPIE.8442E..19H}
M.~{Hazumi}, J.~{Borrill}, Y.~{Chinone}, M.~A. {Dobbs}, H.~{Fuke}, A.~{Ghribi}
  et~al., \emph{{LiteBIRD: a small satellite for the study of B-mode
  polarization and inflation from cosmic background radiation detection}},  in
  \emph{Space Telescopes and Instrumentation 2012: Optical, Infrared, and
  Millimeter Wave}, vol.~8442 of \emph{\procspie}, p.~844219, Sept., 2012,
  \href{https://doi.org/10.1117/12.926743}{DOI}.

\bibitem{2016arXiv161002743A}
K.~N. {Abazajian}, P.~{Adshead}, Z.~{Ahmed}, S.~W. {Allen}, D.~{Alonso}, K.~S.
  {Arnold} et~al., \emph{{CMB-S4 Science Book, First Edition}}, {\emph{ArXiv
  e-prints} (Oct., 2016) }, [\href{https://arxiv.org/abs/1610.02743}{{\ttfamily
  1610.02743}}].

\bibitem{2011arXiv1110.2101K}
B.~{Keating}, S.~{Moyerman}, D.~{Boettger}, J.~{Edwards}, G.~{Fuller},
  F.~{Matsuda} et~al., \emph{{Ultra High Energy Cosmology with POLARBEAR}},
  {\emph{ArXiv e-prints} (Oct., 2011) },
  [\href{https://arxiv.org/abs/1110.2101}{{\ttfamily 1110.2101}}].

\bibitem{quijote2012}
J.~A. {Rubi\~no-Mart\'{\i}n}, R.~{Rebolo}, M.~{Aguiar}, R.~{G\'enova-Santos},
  F.~{G\'omez-Re\~nasco}, J.~M. {Herreros} et~al., \emph{The quijote-cmb
  experiment: studying the polarisation of the galactic and cosmological
  microwave emissions},
  \href{https://doi.org/10.1117/12.926581}{\emph{Proc.SPIE} {\bfseries 8444}
  (2012) 8444 -- 8444 -- 11}.

\bibitem{PhysRevD.55.1830}
M.~Zaldarriaga and U.~c.~v. Seljak, \emph{All-sky analysis of polarization in
  the microwave background},
  \href{https://doi.org/10.1103/PhysRevD.55.1830}{\emph{Phys. Rev. D}
  {\bfseries 55} (Feb, 1997) 1830--1840}.

\bibitem{0004-637X-503-1-1}
M.~Zaldarriaga, \emph{Cosmic microwave background polarization experiments},
  {\emph{The Astrophysical Journal} {\bfseries 503} (1998) 1}.

\bibitem{2015PhRvL.114j1301B}
{BICEP2/Keck Collaboration}, {Planck Collaboration}, P.~A.~R. {Ade},
  N.~{Aghanim}, Z.~{Ahmed}, R.~W. {Aikin} et~al., \emph{{Joint Analysis of
  BICEP2/Keck Array and Planck Data}},
  \href{https://doi.org/10.1103/PhysRevLett.114.101301}{\emph{Physical Review
  Letters} {\bfseries 114} (Mar., 2015) 101301},
  [\href{https://arxiv.org/abs/1502.00612}{{\ttfamily 1502.00612}}].

\bibitem{2003ApJS..148...97B}
C.~L. {Bennett}, R.~S. {Hill}, G.~{Hinshaw}, M.~R. {Nolta}, N.~{Odegard},
  L.~{Page} et~al., \emph{{First-Year Wilkinson Microwave Anisotropy Probe
  (WMAP) Observations: Foreground Emission}},
  \href{https://doi.org/10.1086/377252}{\emph{\apjs} {\bfseries 148} (Sept.,
  2003) 97--117}, [\href{https://arxiv.org/abs/astro-ph/0302208}{{\ttfamily
  astro-ph/0302208}}].

\bibitem{2006PhRvD..74h3002S}
K.~M. {Smith}, \emph{{Pseudo-C$_{ℓ}$ estimators which do not mix E and B
  modes}}, \href{https://doi.org/10.1103/PhysRevD.74.083002}{\emph{\prd}
  {\bfseries 74} (Oct., 2006) 083002},
  [\href{https://arxiv.org/abs/astro-ph/0511629}{{\ttfamily
  astro-ph/0511629}}].

\bibitem{2015MNRAS.452..656V}
M.~{Vidal}, C.~{Dickinson}, R.~D. {Davies} and J.~P. {Leahy}, \emph{{Polarized
  radio filaments outside the Galactic plane}},
  \href{https://doi.org/10.1093/mnras/stv1328}{\emph{\mnras} {\bfseries 452}
  (Sept., 2015) 656--675}, [\href{https://arxiv.org/abs/1410.4438}{{\ttfamily
  1410.4438}}].

\bibitem{2016A&A...594A...9P}
{Planck Collaboration}, R.~{Adam}, P.~A.~R. {Ade}, N.~{Aghanim}, M.~{Arnaud},
  M.~{Ashdown} et~al., \emph{{Planck 2015 results. IX. Diffuse component
  separation: CMB maps}},
  \href{https://doi.org/10.1051/0004-6361/201525936}{\emph{\aap} {\bfseries
  594} (Sept., 2016) A9}, [\href{https://arxiv.org/abs/1502.05956}{{\ttfamily
  1502.05956}}].

\bibitem{2012MNRAS.419.1163B}
S.~{Basak} and J.~{Delabrouille}, \emph{{A needlet internal linear combination
  analysis of WMAP 7-year data: estimation of CMB temperature map and power
  spectrum}},
  \href{https://doi.org/10.1111/j.1365-2966.2011.19770.x}{\emph{\mnras}
  {\bfseries 419} (Jan., 2012) 1163--1175},
  [\href{https://arxiv.org/abs/1106.5383}{{\ttfamily 1106.5383}}].

\bibitem{2002ApJ...567....2H}
E.~{Hivon}, K.~M. {G{\'o}rski}, C.~B. {Netterfield}, B.~P. {Crill}, S.~{Prunet}
  and F.~{Hansen}, \emph{{MASTER of the Cosmic Microwave Background Anisotropy
  Power Spectrum: A Fast Method for Statistical Analysis of Large and Complex
  Cosmic Microwave Background Data Sets}},
  \href{https://doi.org/10.1086/338126}{\emph{\apj} {\bfseries 567} (Mar.,
  2002) 2--17}, [\href{https://arxiv.org/abs/astro-ph/0105302}{{\ttfamily
  astro-ph/0105302}}].

\bibitem{2010A&A...519A.104K}
J.~{Kim} and P.~{Naselsky}, \emph{{E/B decomposition of CMB polarization
  pattern of incomplete sky: a pixel space approach}},
  \href{https://doi.org/10.1051/0004-6361/201014739}{\emph{\aap} {\bfseries
  519} (Sept., 2010) A104}, [\href{https://arxiv.org/abs/1003.2911}{{\ttfamily
  1003.2911}}].

\bibitem{PhysRevD.82.023001}
W.~Zhao and D.~Baskaran, \emph{Separating $e$ and $b$ types of polarization on
  an incomplete sky},
  \href{https://doi.org/10.1103/PhysRevD.82.023001}{\emph{Phys. Rev. D}
  {\bfseries 82} (Jul, 2010) 023001}.

\bibitem{2017PhRvD..96d3523B}
E.~F. {Bunn} and B.~{Wandelt}, \emph{{Pure E and B polarization maps via Wiener
  filtering}}, \href{https://doi.org/10.1103/PhysRevD.96.043523}{\emph{\prd}
  {\bfseries 96} (Aug., 2017) 043523},
  [\href{https://arxiv.org/abs/1610.03345}{{\ttfamily 1610.03345}}].

\bibitem{2018arXiv180105358K}
D.~{Kodi Ramanah}, G.~{Lavaux} and B.~D. {Wandelt}, \emph{{Optimal and fast E/B
  separation with a dual messenger field}}, {\emph{ArXiv e-prints} (Jan., 2018)
  }, [\href{https://arxiv.org/abs/1801.05358}{{\ttfamily 1801.05358}}].

\bibitem{2018JCAP...05..059L}
H.~{Liu}, J.~{Creswell} and P.~{Naselsky}, \emph{{E and B families of the
  Stokes parameters in the polarized synchrotron and thermal dust
  foregrounds}},
  \href{https://doi.org/10.1088/1475-7516/2018/05/059}{\emph{\jcap} {\bfseries
  5} (May, 2018) 059}, [\href{https://arxiv.org/abs/1804.10382}{{\ttfamily
  1804.10382}}].

\bibitem{2018A&A...617A..90L}
H.~{Liu}, \emph{{Fingerprint of Galactic Loop I on polarized microwave
  foregrounds}}, \href{https://doi.org/10.1051/0004-6361/201833471}{\emph{\aap}
  {\bfseries 617} (Sept., 2018) A90},
  [\href{https://arxiv.org/abs/1806.06532}{{\ttfamily 1806.06532}}].

\bibitem{WMAPdata:online}
{The WMAP data release}. \url{https://lambda.gsfc.nasa.gov/product/map/dr5/ },
  2013.

\bibitem{Planckdata:online}
{The Planck data release}. \url{https://www.cosmos.esa.int/web/planck/pla },
  2018.

\bibitem{2005ApJ...622..759G}
K.~M. {G{\'o}rski}, E.~{Hivon}, A.~J. {Banday}, B.~D. {Wandelt}, F.~K.
  {Hansen}, M.~{Reinecke} et~al., \emph{{HEALPix: A Framework for
  High-Resolution Discretization and Fast Analysis of Data Distributed on the
  Sphere}}, \href{https://doi.org/10.1086/427976}{\emph{\apj} {\bfseries 622}
  (Apr., 2005) 759--771},
  [\href{https://arxiv.org/abs/astro-ph/0409513}{{\ttfamily
  astro-ph/0409513}}].

\end{thebibliography}

\providecommand{\href}[2]{#2}\begingroup\raggedright\endgroup

\end{document}